\documentclass[journal,transmag]{IEEEtran}

\ifCLASSINFOpdf
\usepackage[pdftex]{graphicx}
\else
\fi
\usepackage{amsmath}
\usepackage{hyperref}
\begin{document}
\title{Decoding Political Polarization in Social Media Interactions}


\author{\IEEEauthorblockN{Giulio Pecile\IEEEauthorrefmark{1},
Niccolò Di Marco\IEEEauthorrefmark{1},
Matteo Cinelli\IEEEauthorrefmark{1}, 
Walter Quattrociocchi\IEEEauthorrefmark{1}}
\IEEEauthorblockA{\IEEEauthorrefmark{1}Sapienza University of Rome, Italy}%
\thanks{%Manuscript received December 1, 2012; revised August 26, 2015. 
Corresponding author: W. Quattrociocchi (email: walter.quattrociocchi@uniroma1.it).}}%

% \markboth{Journal of \LaTeX\ Class Files,~Vol.~14, No.~8, August~2015}%
% {Shell \MakeLowercase{\textit{et al.}}: Bare Demo of IEEEtran.cls for IEEE Transactions on Magnetics Journals}

\IEEEtitleabstractindextext{%
\begin{abstract}
Social media platforms significantly influence ideological divisions by enabling users to select information that aligns with their beliefs and avoid opposing viewpoints. Analyzing approximately 47 million Facebook posts, this study investigates the interactions of around 170 million users with news pages, revealing distinct patterns based on political orientations. While users generally prefer content that reflects their political biases, the extent of engagement varies even among individuals with similar ideological leanings. Specifically, political biases heavily influence commenting behaviors, particularly among users leaning towards the center-left and the right. Conversely, the 'likes' from center-left and centrist users are more indicative of their political affiliations. This research illuminates the complex relationship between social media behavior and political polarization, offering new insights into the manifestation of ideological divisions online.
\end{abstract}

}

\maketitle
\IEEEdisplaynontitleabstractindextext
\IEEEpeerreviewmaketitle

\section{Introduction}
\IEEEPARstart{T}he World Wide Web has vastly increased information accessibility, transforming how information is distributed and consumed globally. This evolution has revolutionized communication methods, erasing geographical and temporal constraints. Over recent decades, expanding social networks and shifting public discourse to digital platforms have led scholars to explore how users interact with information and form behavioral clusters, often challenging traditional expectations.
It is well-documented that online users tend to engage with information that confirms their pre-existing beliefs, typically ignoring conflicting viewpoints \cite{Bakshy.2015,Zollo.2017}. Such behaviors create echo chambers, where like-minded individuals reinforce each other's views \cite{Zollo.2015, Garimella.2018, DelVicario.2016}. Such configurations, alongside growing polarization, manifest with varying intensity across different social media platforms \cite{Cinelli.2021}, suggesting that the users' base as well as platform designs and algorithms, which aim to maximize engagement, play a significant role in shaping these social dynamics.

Existing literature \cite{Garrett.2009} suggests that although individuals generally prefer political news that aligns with their pre-existing beliefs, they do not entirely ignore opposing viewpoints. Conversely, Guess \cite{Guess.2016} finds that most users gravitate towards centrist news outlets despite a small, highly engaged group favoring partisan sources. These patterns indicate a notable convergence in the information available to the broader public. However, the ability of individuals to completely shield themselves from conflicting information is limited due to the minimal control they have over the content presented to them and the dynamics of information spread on social media platforms. Notably, 'weak ties'—connections between loosely affiliated individuals like acquaintances or distant relatives—are crucial in spreading new information across social networks \cite{Bakshy.2012}.

Notably, most content users interact with on social media is not actively sought but is delivered through a network algorithm tailored to maximize user engagement. This characteristic is pivotal as it may foster unrecognized behavioral patterns, complicating efforts to measure such effects. Due to concerns over potential manipulation by malicious actors, social networks often refrain from disclosing the specifics of these algorithms. Sunstein, in his book \textit{Republic: Divided Democracy in the Age of Social Media} \cite{sunstein.2018}, discusses the \textit{Age of the Algorithm}, where users lack control over their news consumption, inadvertently contributing to the formation of echo chambers.

Not all users exhibit polarization similarly. However, their behaviors show significant nuances. Zaller observes that politically engaged citizens are particularly receptive to messages that align with their beliefs \cite{zaller.1992}. Similarly, Taber and Lodge find that highly partisan users are more likely to embrace supporting arguments while dismissing contrary ones without question \cite{taber.2006}. This suggests that increasing polarization complicates efforts to mitigate it, and using counterfactuals might even be counterproductive \cite{Zollo.2017}. The debate over which political group is more prone to selective exposure and biased information processing remains unresolved. Some studies indicate that conservatives are more likely to engage in such behaviors \cite{lau.2006, nyhan.2010, Nam.2013}, while others present contradictory findings \cite{munro.2002, iyengar.2012, nisbet.2015}. Furthermore, there is no consensus on cross-party discussions; Barberá suggests that liberals are more involved in cross-party interactions \cite{barbera.2015}, whereas Wu argues that conservatives are more likely to engage in such discussions \cite{siqi.2021}.

In this study, we explore the phenomenon of selective exposure by analyzing how about 170 million Facebook users interact with approximately 47 million posts from news agencies with varying political leanings. We specifically measure the intensity of selective exposure and assess whether users prefer specific ideologies or news agencies. The analysis is structured around several distinct scenarios of user selectivity:

\begin{itemize}
 \item users who are not selective at all; 
\item users who are selective both in terms of pages viewed and the political leaning of the pages; 
\item users who are selective only in terms of political leaning; 
\item users who are selective in terms of pages viewed.
 \end{itemize}

By categorizing user behavior into these scenarios, we aim to uncover the extent and nature of selective exposure, identifying whether it is more pronounced towards particular ideologies or news providers.

We find that all users exhibit strong selectivity in terms of political leaning, a phenomenon explained mainly by a marked preference for specific pages \cite{Cinelli.2020, Schmidt.2017}. However, this preference does not fully account for the leaning-based selective exposure observed in specific user groups.
These results are crucial for understanding the primary drivers of selective exposure and the resultant polarization within online communities. Our framework enables an examination of the role political affiliation plays in the selectivity exhibited by users.
The structure of this paper is organized as follows: Initially, we discuss related works that explore the concepts of polarization, the echo chamber hypothesis, and their interactions with social media. We then describe the theoretical methods used in our analysis, including entropy as a proxy for measuring selective exposure and two randomization strategies to evaluate the robustness of our findings. Following this, we detail the patterns of user activity concentration and demonstrate how this concentration aligns predominantly with ideological page biases. The paper concludes by presenting evidence of leaning-driven selectivity, predominantly among users who follow pages with right and center-left biases.

\section{Related works}

\subsection{Polarization and Echo Chambers}
The prevailing hypothesis suggests that echo chambers on social media significantly influence social dynamics by amplifying similar opinions and minimizing opposing ones, fostering homophily and polarization \cite{DelVicario.2016}. This effect is often exacerbated by the emergence of intolerance in online discussions \cite{del2016echo,bessi.2016}, which further polarizes interactions \cite{bessi2015viral, del2017modeling,cheng.2017, Saveski.2021}. Numerous studies employing diverse algorithms and methodologies substantiate these observations \cite{valensise2023drivers,baumann2020modeling,cinus2022effect,terren2021echo}, including Garimella's work \cite{Garimella.2018}, which uses the Walktrap Controversy metric to measure segregation in discussion networks on polarizing topics. Additionally, Salloum \cite{Salloum.2022} investigates how network size, edge count, and degree distribution influence polarization scores, proposing methods to mitigate these biases.
Global events such as Brexit \cite{DELVICARIO20176}, vaccine debates \cite{schmidt2018polarization}, and climate change discussions \cite{Falkenberg.2021} frequently act as catalysts for the formation of echo chambers. The political bias of news agencies covering these events significantly contributes to these dynamics, as studies indicate that highly polarized user clusters often form around specific news pages \cite{Schmidt.2018, Cinelli.2020}, distinguished by their narrative and selective biases \cite{Galeazzi.2023}.
It is crucial to differentiate between affective and ideological polarization. Affective polarization, or psychological polarization \cite{settle.2018}, occurs when opposing groups harbor feelings of dislike and distrust towards each other. In contrast, ideological polarization involves differing views that do not necessarily include moral judgments of the opposing group. Affective polarization is particularly noted in the partisan divide between Democrats and Republicans in the United States \cite{mason.2018} and is increasingly prevalent on social media. This rise is attributed to the ease of identifying a user’s political leanings, which complicates civil cross-party dialogue, reinforces social and political identities, and promotes negative stereotyping of opposing groups.

\subsection{The interplay between social media and polarization}
While there is broad consensus that social networks often serve as arenas for polarization and echo chambers, the precise mechanisms through which these platforms influence these phenomena are still being explored \cite{Cinelli.2021}. Concerns about the role of social media and search engines in filtering news, potentially exacerbating polarization, remain significant \cite{sunstein.2018, pariser.2012}. Although highly polarized users tend to form homogeneous clusters, such filtering effects might be mitigated by systems like those proposed by Garimella \cite{garimella.2017}. Conversely, studies indicate that exposure to contrarian news sources may actually increase political polarization \cite{rathje2021out,bail.2018}, and introducing users who believe in conspiracy theories to fact-checkers could be ineffective or even counterproductive \cite{Zollo.2017,cinelli2022conspiracy}.
The impact of algorithmically tailored feeds on polarization has also been critically examined \cite{gonzalesbailon.2023, guess.2023_1, guess.2023_2, nyhan.2023}. These studies assess the interactions between algorithmic recommendations and societal forces, comparing their effects to those produced by non-tailored feeds. The findings suggest that personalized feeds may not significantly increase polarization, pointing to other factors as potential drivers behind the observed polarization on these platforms.

\section{Materials and Methods}
\subsection{Entropy and selective exposure\label{mmsubentropy}} 

To fully measure the phenomenon of selective exposure defined as \textit{a tendency for people both consciously and unconsciously to seek out material that supports their existing attitudes and opinions and to actively avoid material that challenges their views} \cite{selectiveexposure}, one would need the full digital trace of a user and the reasons that motivated the user to interact with one page instead of another. In its absence, we use the entropy of the interactions as a proxy. We recall that Shannon Entropy is commonly used in information theory to measure the concentration of a distribution. In our framework, users with low entropy are the most selective, while those interacting with different bias levels uniformly have high entropy.

In particular, here we also employ some of its properties that arise when considering sub-partitions.

Consider a set universe $U$ and a partition $\sigma = \{s^1, \ldots, s^n\}$ of it. We denote $\lvert s^i \rvert \equiv c_i$. 
Suppose that $\rho$ is a partition of $\sigma$, i.e. if for each $r \in \rho, r \subseteq s_i \in \sigma$ for exactly one $i$. Then, each set of $\rho$ can be written as:

$$
\rho = \{s_1 ^1 ,\ldots,s_{c_1}^1, \ldots, s_1 ^i,\ldots,s_{c_i}^i, s_1 ^n ,\ldots,s_{c_n}^n\},
$$

i.e. $s_j ^i$ its the $j-$th subset of $s^i$.

Now, consider a random variable $X_{\rho}$ having image in $\rho$. Obviously, $X_{\rho}$ can be naturally extended to $X_{\sigma}$ having image in $\sigma$. We define $p(X_\rho \in s_j ^i) \equiv p_j ^i$. It follows,  $p(X_{\rho} \in s^i) = p(X_{\sigma} \in s^i) = \sum_{j = 1}^{c_i} p_j ^i \equiv p^i$. 
We have that:

$$
H(X_\rho) = - \sum_i \sum_j p_j ^i \log p_j^i =
$$

$$
= - \sum_i \sum_j \left( p_j ^i \log(p^i) + p_j ^i \log\frac{p_j^i}{p^i} \right) =
$$

$$
= - \sum_i p^i log(p^i) - \sum_i p^i \sum_j \bar{p}_j^i \log \bar{p}_j^i,
$$

where $\bar{p}_j ^i = \frac{p_j^i}{p^i}$. In a more compact form, we have obtained:

\begin{equation}\label{eq:entropy}
    H(X_\rho) = H(X_\sigma) + \sum_i p^i H(X_\rho  | X_{\sigma} \in s^i).
\end{equation}

In this work, we consider interactions over pages having a certain political leaning. Therefore, given the general set of spaces, $\sigma$ will be the partition induced by their political leaning, while $\rho$ will be simply the partition in which each page is considered alone.

Having reached Equation \eqref{eq:entropy}, it is immediate to find the theoretical minimum and maximum entropy of the finer partition $\rho$ (i.e. the pages) given the interaction over partition $\sigma$ (i.e. the leanings). 

This follows from the observation that the $H(X_\sigma)$ and $p^i$ terms are fixed, and the terms $H(X_\rho | X_\sigma \in s^i)$ are all independent of one another and thus can be minimized (maximized) independently.  
The minimum is clearly found when $H(X_\rho | X_\sigma \in s^i) = 0\; \forall i$ i.e. when the activity of the user is concentrated in at most one page for every $s^i$ (i.e. leaning).

On the other hand, we recall that the maximum entropy is reached by a uniform distribution.
Since we consider interactions, which are discrete and often not large, it is not always possible to obtain an exact uniform distribution. Therefore, we compute the maximum possible value of entropy with the below criteria. 

Consider a user $u$ with $n$ interactions in $c_i$ pages from $s^i$. The maximum entropy is found when $r$ pages receive $q+1$ interactions and $c_i-r$ pages receive $q$ interactions, where \( q \) and $r$ are the quotient and remainder of the integer division of $\frac{n}{c_i}$. Note that, if $n\leq c_i$, the maximum entropy is simply $\log{n}$.

Finally, we note that in every $s^i$ the minimum and maximum entropies are the same only when there is only one interaction, in which case $H(X_\rho | X_\sigma \in s^i) = 0$.

\subsection{Strong and weak randomization \label{mmrandomizations}} 
\begin{figure}[ht]
  \centering
  \includegraphics[width = \linewidth]{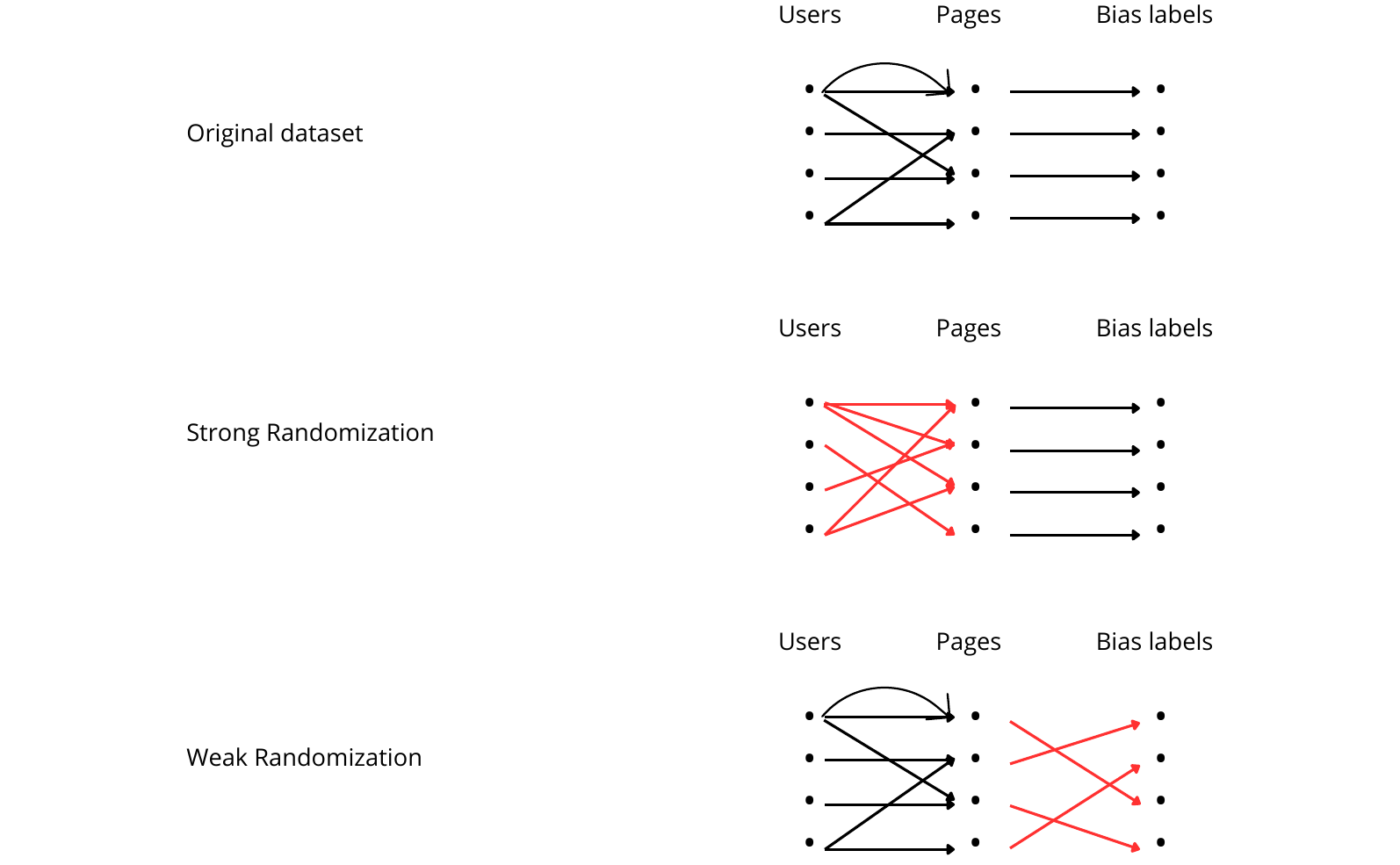}
  \caption{In the tripartite representation of the interactions, the strong randomization process affects the user-pages interactions, while the weak one affects the page-bias affiliations.}
  \label{fig:randomizations}
\end{figure}

In our analysis, we compare the actual interaction patterns to those resulting from two randomization processes: a stronger one, where each interaction between users and pages is randomized, and a weaker one, where the Bias labels of the pages are randomized. In both cases, the number of interactions made by each user and received by each page remains the same, and the number of pages for each bias label remains unchanged. Furthermore, we notice that in the strong randomization process, the users distribute their activity uniformly across pages as the original multiple interactions between the same user and the same page are now spread between multiple pages. This uniformity of interactions implies that the result of a strong randomization process would be unaffected by a further weak randomization. Thus, the strong randomization process implies the weaker one. In Figure \ref{fig:randomizations}, the details of the randomization processes are visually explained. Note that, in the bipartite representation, each edge indicate an interaction between a user and a page and multiple edges are allowed.

\section{Results}
\begin{figure}[ht]
  \centering
  \includegraphics[width = \linewidth]{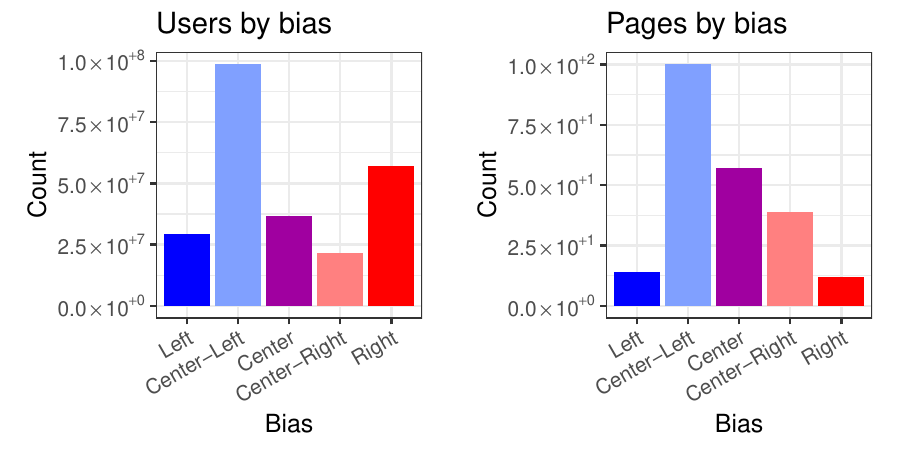}
  \caption{Number of Facebook pages and users grouped by political affiliation.}
  \label{fig:leaningdistributions}
\end{figure}

In our analysis, we use the same dataset as two previous papers \cite{Cinelli.2020,Cinelli.2021}, which comprise roughly 266 million comments and 1.5 billion likes from about 170 million users on 47 million Facebook posts by 222 pages of news agencies. Those news agencies have a Bias score, provided by Media Bias/Fact Check \cite{MBFC}, with five possible leanings ranging from left to right. We infer the political bias of users using the mode of the leanings of their interactions. The distribution of pages and users by political leaning is shown in Figure \ref{fig:leaningdistributions}, where we see that while most pages and users are identified as center-left leaning, there are in proportion many more right-wing users than pages. 

\subsection{Measuring Selective exposure}
\begin{figure}[!ht]
  %\centering
  \includegraphics[width = \linewidth]{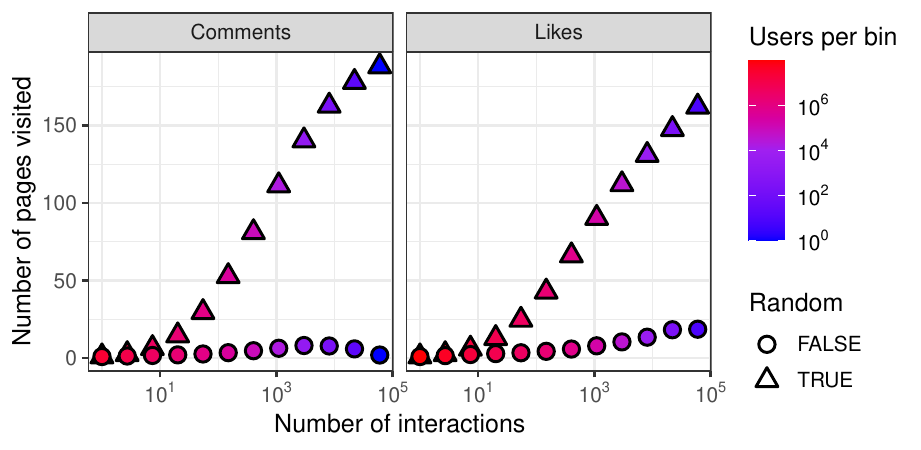}
  \caption{Number of pages among which users divide their attention.}
  \label{fig:act_p_page}
\end{figure}

We first explore patterns of activity concentration by grouping users by the number of their likes and comments.
We create $12$ logarithmic bins and, for each bin, we compute the average number of pages they interact with. We then replicate this analysis on the strongly randomized dataset (see Materials and Methods for further details). Figure \ref{fig:act_p_page} compares the two scenarios.

The results show that the real behavior of users is much more concentrated than the randomized case. Note that, in this latter case, the probability that a user interacts with a page is solely based on the page's popularity i.e. the number of users interacting with it.
While this on itself does not imply that there is a specific tendency for users to prefer content that is congruent with their specific view, it already suggests that users' activity is strongly dependent on factors that transcend the simple popularity of the page, as in the randomization case. 

\begin{figure}[!ht]
  \centering
  \includegraphics[width = \linewidth]{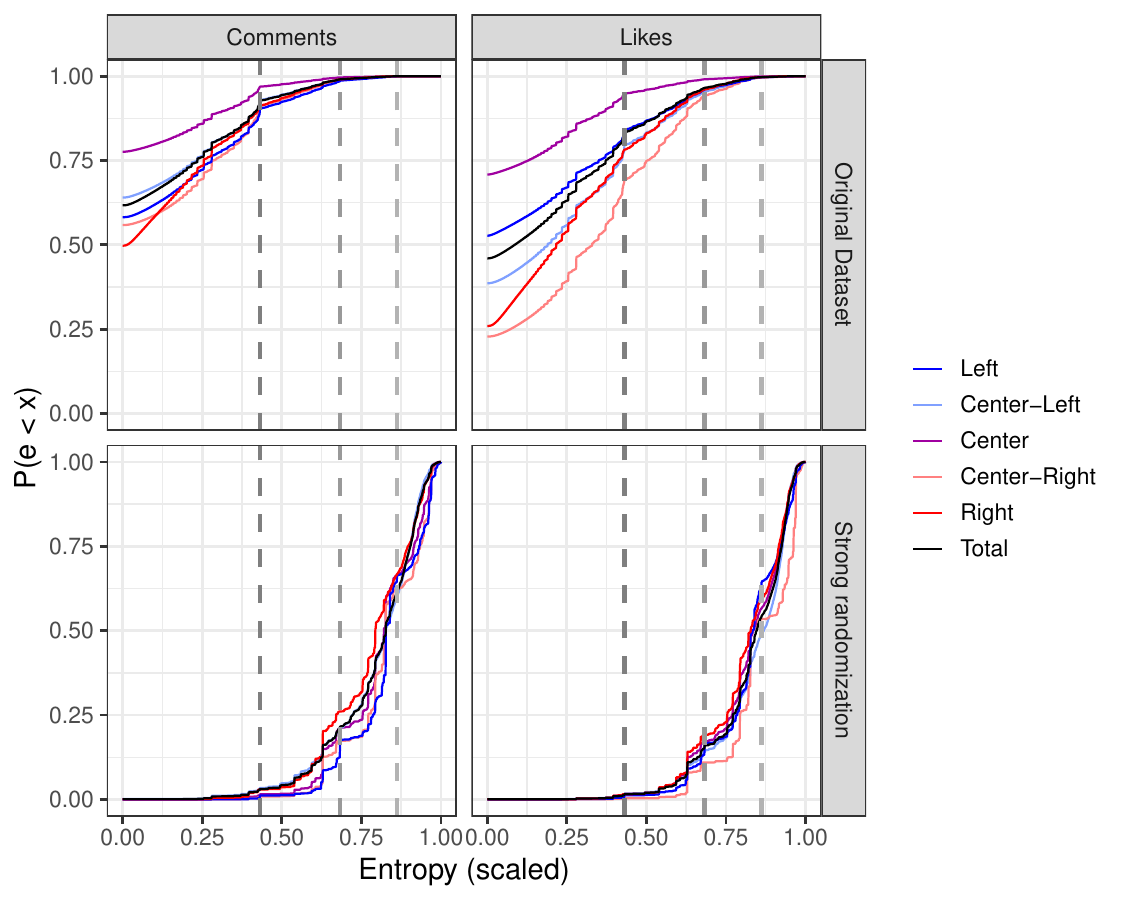}
  \caption{Distributions of bias entropy of users compared with the strongly randomized scenario.}
  \label{fig:strongentropy}
\end{figure}

To understand if political leaning could be one of these factors, we group pages by their bias level and, for each user, we evaluate the Shannon entropy of their distribution of interactions left in each political class. We use this metric because it clearly indicates how concentrated or spread out each user's activity is. We scale the entropy values to a $[0, 1]$ interval by dividing by $\log{5}$, the theoretical maximum entropy value for 5 classes. To ensure that the observed selectivity is not due to low activity, we consider only users who have made more than 5 comments (likes) in this and all further analyses. The empirical cumulative distribution function (eCDF) of the values of Shannon's entropy for all groups of users is presented in Figure \ref{fig:strongentropy}. 
Additionally, we highlight the entropy levels corresponding to a user interacting evenly with two, three, and four different leanings using grey lines.

Clearly, the leaning classification of users is most reliable for those with entropy at the left of the first line, since they comment very heterogeneously.  All groups of users display a strong level of selectivity, with center-leaning users being the most selective. We observe that, in all groups, at least 50\% of the users concentrated their commenting activity on pages of the same political leaning. This trend is also present, albeit less prominent, with likes, with center-right-leaning users being less selective than other groups of users. We compare this situation with the strong randomization process, that describes users interacting with pages based solely on popularity. 

Interestingly, we observe a substantial difference with the randomized scenario, suggesting again that even the least selective users choose content according to a criterion compatible with political leaning. 

\subsection{Page and Bias Selectivity}
In the previous section we have shown that users display a strong level of selectivity compatible with an ideological classification of the pages. However, this can only partially capture users' preferences in their activity. If users base their interactions solely on the political leaning of a page, they will treat all pages with the same leaning similarly. On the other hand, if users are selective about specific pages, their interactions within each bias level will reflect that page-driven selectivity.
We observe that since pages can be thought of as a sub-partition of the bias levels, the values of the two entropies (the one calculated by considering the interaction on pages and bias levels) are actually dependent, as explained in Materials and Methods. In particular, we find that the entropy calculated on pages is equal to the entropy calculated on the bias levels summed with a weighted average of the entropies of the pages inside each bias level. Since each term of this average is independent, they can be trivially minimized and maximized. 
As explained in Materials and Methods, for each user, we find the theoretical interval $I(u) = [m(u), M(u)]$ in which the actual page entropy $H_p(u)$ can be found and scale the result using

$$
    x(u) = \frac{H_p - m}{M-m}.
$$

Doing that, users who make at most one interaction per bias level are removed as the theoretical minimum and maximum coincide. This phenomenon is desirable, as it is impossible to decide if further selectivity is motivated by a preference for specific pages. We also recall that the analysis is performed only on users who have made at least five comments (likes) to ensure that the selectivity measured cannot be attributed to low activity. Table \ref{tab:pavegivenbiasc} reports the summary characteristics of the $x(u)$ values distribution. Interestingly, we can observe that users tend to be very selective regarding pages, often concentrating their activity on one page per bias level.

\begin{table}
\caption{Quartiles of the $x$ statistics describing inter-bias selectivity.}
    \centering
    \begin{tabular}{c|c|c|c|c|c}
         & Min & 1st Quartile & Median & 3rd Quantile & Max  \\
         \hline
        Likes & 0 & 0 & 0 & 0.2108 & 1  \\
        \hline
        Comments & 0 & 0 & 0 & 0.09045 & 1\\
    \end{tabular}
    \label{tab:pavegivenbiasc}
\end{table}

\subsection{Bias and Page Selectivity}
In the previous sections, we have observed that, although users exhibit a selective behavior that is compatible with political segregation, it is not the only explaining driver of user activity. As users interact with -relatively- few pages, many alternative divisions could be compatible with the observed phenomenon, and thus, it is necessary to assess the divergence between the real selective exposure expressed by the users and a benchmark value found by grouping pages randomly. To accomplish this, we perform the weak randomization process described in Materials and Methods to account for measurements of bias selectivity that may arise from the simple page selectivity or other sorts of selective behaviors (for instance, locality-driven selectivity). We notice that the users interacting with only one page will naturally interact with only one bias class, regardless of the randomization. For those users, deciding if a preference in political leaning, locality, or any arbitrary classification of pages drives their choice of activity is impossible. For this reason, we set those users aside and only compute the distributions of users who interact with at least two pages and have a minimum of five interactions. 
Upon performing the weak randomization process, we note that since the user-page relations are unchanged, the page entropy of each user remains unaffected by this particular randomization process. This allows us to measure the bias selectivity while accounting for the strong patterns of page selectivity measured previously. 

We perform a Monte Carlo experiment where each simulation is obtained using the weak randomization process. To ensure the manageability of the procedure, we used a random sample of the dataset (slightly over 2\% of the users).  By averaging over multiple groupings of pages, we obtain a benchmark value for the bias entropy of the users, which we can then compare to the actual bias entropy. If the political leaning determines a further selectivity, the distribution of the original dataset will have higher entropy values, and, especially at the left of the first line, the entropy eCDF will have higher values. If, on the contrary, the two distributions are similar or the weak randomization process produces higher entropy values, it means that the political leaning of the pages does not drive the selectivity of the users.

\begin{figure*}[!ht]
  \centering
  \includegraphics[width = \linewidth]{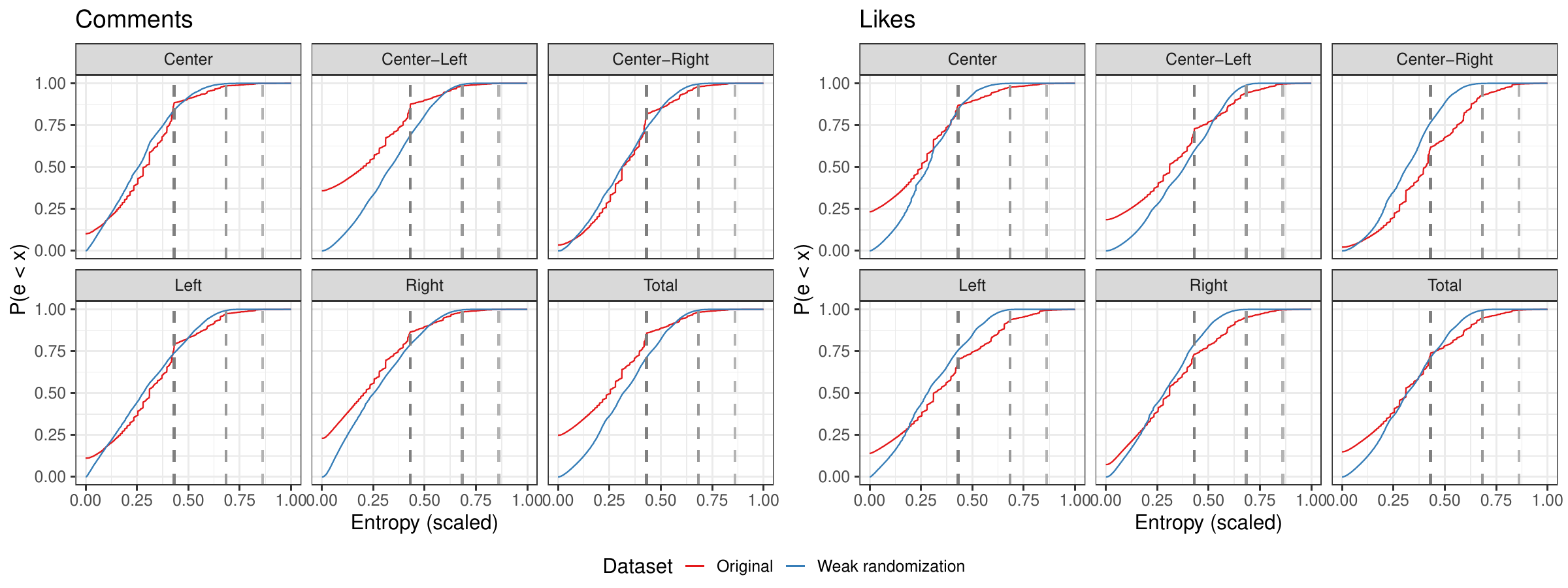}
  \caption{Distribution of users' bias entropy compared with the benchmark values obtained via weak randomization.}
  \label{fig:biasgivenpagel}
\end{figure*}

Figure \ref{fig:biasgivenpagel} compares the real and randomized eCDF distributions and Table \ref{tab:kldiv} summarizes the Kullback-Leibler divergence between the distributions.

We observe that users primarily commenting on center and center-right-leaning pages align closely with the randomized distribution, suggesting that the political leaning of these pages aligns with their usual consumption patterns. In contrast, users interacting with center-left-leaning pages show the most significant divergence from the randomized model, indicating that political leanings heavily influence their news consumption habits. This is evident when comparing these findings with those illustrated in Figure \ref{fig:strongentropy}, where users engaging with only one page are included. Notably, the exclusion of such users significantly reduces the number of those with zero entropy. Center-leaning users, who are the most selective, are most impacted by this exclusion. This suggests that less politically engaged users, who typically follow only one page, tend to prefer center-leaning pages. Conversely, users with more explicit political preferences display greater selectivity in their interactions. For likes, the patterns slightly differ. Center-leaning users clearly exhibit bias-driven engagement, a trend that is less pronounced among right-leaning users and even less so among center-right-leaning ones. 

In particular, center-right-leaning users are less selective than their randomized counterparts. Left- and center-left-leaning users display prominent levels of divergence.

\begin{table*}
    \centering
    \caption{The Kullback-Leibler divergence between the distribution of bias-entropies and the benchmark obtained via the weak randomization.}
    \begin{tabular}{c|c|c|c|c|c|c}
         & Left & Center-Left & Center & Center-Right & Right & Total\\
         \hline
         Likes & 1.029326 & 1.110854  & 0.948667  & 0.8790362  & 0.6452442  & 1.002449 \\
         \hline
         Comments & 0.5204064  & 1.317479  & 0.4297106  & 0.480125  & 0.6581254  & 1.014246  \\
    \end{tabular}
    
    \label{tab:kldiv}
\end{table*}

\section{Conclusion}
In this paper, we conduct a comprehensive analysis of the phenomenon of selective exposure on social media platforms. Initially, we observe that user activity predominantly concentrates on a limited number of pages. As user engagement increases, these pages quickly become saturated, suggesting that user interactions are focused despite the availability of vast content. Employing Shannon Entropy to explore the homogeneity of user behavior, we identify a strong preference for content that aligns with users' pre-existing political views. This finding supports the hypothesis that social media serves as an echo chamber, amplifying and reinforcing similar viewpoints.
Further analysis reveals that user engagement is not uniform across pages with similar political leanings. Users prefer specific news sources within political categories, indicating a more nuanced approach to selective exposure that includes specific sources resonating deeply with individual users. This observation prompted developing and applying a novel methodology designed to dissect and analyze the reasons behind such selective behaviors. Our findings indicate that political congruence is a more significant driver of user behavior than random selection, with effects particularly pronounced among users leaning towards the center-left.
Our work establishes a robust framework for analyzing and comparing the mechanisms of selective exposure across various user groups on social media. This framework enhances our understanding of why users gravitate towards certain content and improves our ability to predict page-level selectivity based on political bias. Additionally, it helps identify which user groups are most susceptible to selective exposure, shedding light on how echo chambers form and persist.
However, this study has limitations. It does not account for the tone or nature of interactions—whether users support, criticize, or comment on content in a neutral or hostile manner. Despite this, we are confident in the robustness of our findings. We observe that negative or hostile interactions, though present, do not significantly alter the overall patterns of selective exposure.

\bibliography{main}

% or


\appendices
%% you can choose not to have a title for an appendix
%% if you want by leaving the argument blank
%\section{}

\section*{Acknowledgment}

The work is supported by IRIS Infodemic Coalition (UK government, grant no. SCH-00001-3391), 
SERICS (PE00000014) under the NRRP MUR program funded by the European Union - NextGenerationEU, project CRESP from the Italian Ministry of Health under the program CCM 2022, PON project “Ricerca e Innovazione” 2014-2020, and PRIN Project MUSMA for Italian Ministry of University and Research (MUR) through the PRIN 2022.

\ifCLASSOPTIONcaptionsoff
  \newpage
\fi



\bibliographystyle{IEEEtran}
\end{document}